\documentclass[prl,aps,amssymb,twocolumn,superscriptaddress,showpacs]{revtex4}
\usepackage{amsmath}
\usepackage{amssymb}
\usepackage{color}
\usepackage{graphics}
\usepackage{graphicx}
\usepackage{enumerate}
\usepackage{latexsym}
\usepackage{dcolumn}
\usepackage{youngtab}
\makeatletter
\renewcommand\normalsize{%
   \@setfontsize\normalsize\@xpt\@xiipt
   \abovedisplayskip 1\p@ \@plus2\p@ \@minus5\p@
   \abovedisplayshortskip \z@ \@plus3\p@
   \belowdisplayshortskip 6\p@ \@plus3\p@ \@minus3\p@
   \belowdisplayskip \abovedisplayskip
   \let\@listi\@listI}
\makeatother

\def\k{\mbox{\boldmath$\kappa$}}

\def \be {\begin{equation}}
\def \ee {\end{equation}}
\def \bea {\begin{eqnarray}}
\def \eea {\end{eqnarray}}
\def \beaa {\begin{eqnarray*}}
\def \eeaa {\end{eqnarray*}}
\def \nn  {\nonumber}

\def\bra#1{\langle{#1}|}
\def\ket#1{|{#1}\rangle}

\begin{document}

\title {Calogero-Sutherland model in interacting fermion picture and explicit construction of Jack states}
\author{ Jian-Feng Wu}
\email{wujf@itp.ac.cn}
\author{ Ming Yu}
\email{yum@itp.ac.cn}
 \affiliation{Institute of Theoretical Physics, Chinese Academy of Sciences, Beijing, China, 100190}
\begin{abstract}
The 40-year-old Calogero-Sutherland (CS) model remains a source of inspirations for understanding 1d interacting fermions.
At $\beta=1, \text{or }0$, the CS model describes a free non-relativistic
fermion, or boson theory, while for generic $\beta$, the system can be interpreted either as interacting fermions or bosons, or free anyons depending on the context. However,
we shall show in this letter that  the fermionic picture is advantageous in diagonalizing the CS Hamiltonian. Comparing to the previously known multi-integral representation or the Dunkl operator formalism for the CS wave functions, our method depends on the (upper or lower) triangular nature of the fermion interaction, which is resolved in perturbation theory of the second quantized form.
The eigenstate is constructed from a multiplet of unperturbed states and the perturbation is of finite order. The full construction is a similarity transformation from the free fermion theory, in the same spirit as the Landau Fermi liquid theory and the 1d Luttinger liquid theory. That means quasi-particles or anyons can be represented in terms of free fermion modes (or bosonic modes via bosonization). 
The method is applicable to other  (higher than one space dimension)  systems for which the adiabatic theorem applies.
\end{abstract}
\pacs{71.10.Pm, 11.25.Hf, 02.30.Ik} \maketitle
In this letter we shall propose an explicit formula for solving a class of Hamiltonian eigenequation and work out the explicit construction of the Jack states for the CS model\cite{Calogero:1969,Sutherland:1971}  as a specific example. Comparing to the previously known multi-integral representation\cite{Awata:1995ky, Wu:2011cya} or the Dunkl operator formalism\cite{Lapointe(1995)} for the CS wave functions, our method depends on the (upper or lower) triangular nature of the fermion interaction, which is resolved in perturbation theory of the second quantized form. The similar method has also been used in a different context on the explicit construction of the AFLT states\cite{Shou:2011nu}. 
The general statement on the Hamiltonian system to which our construction applies
is as following:
An interacting Hamiltonian system (with Hamiltonian $H$) or its equivalent class by a similarity transformation, is exactly solvable through finite order perturbation if its matrix form is (upper or lower) triangular which will be abbreviated just as triangular hereafter. In practice, we choose as the basis vectors the already solved eigenstates for an ``unperturbed''  Hamiltonian $H^{(0)}$, $H^{(0)}|E^{(0)}\rangle_0 =E^{(0)}|E^{(0)}\rangle_0$.  Although not necessarily, one can choose the free theory as the unperturbed system, in which the interaction is turned off. $H$ will contain perturbations away from $H^{(0)}$,  $H=H^{(0)} +H^{(I)}$. The crucial point is that we shall assume, although not always  guaranteed to be so, that $H^{(I)}$ can be decomposed into $H^{(I)}=H^{\parallel}+H^{\perp}$. Such decomposition makes our method differing substantially from the others' triangulating the Hamiltonian\cite{Lapointe(1995), Sutherland:Lecture}.   Here, $H^{\parallel}$ is diagonal with diagonal entry $E^{(I)}$ and $H^{\perp}$ is strictly triangular in the basis of the $H^{(0)}$ eigenstates. By ``strictly triangular'' we mean the triangular matrix with zero diagonal entries. 
The hermiticity of $H$, if lost, will be restored by the inverse-similarity transformation. We may write $H^d=H^{(0)}+H^{\parallel},\,   E=E^{(0)}+E^{(I)},\, |E\rangle = R(E)|E^{(0)}\rangle_0$.  
Then the energy eigenequation $H|E\rangle = E|E\rangle$ is solved with the following solution,
$
R(E) = \bigl(1-(E-H^d)^{-1} H^{\perp}\bigr)^{-1}= \sum_{n=0}^{\infty}\bigl((E-H^d )^{-1}H^{\perp}\bigr)^n
$.  
This can be checked by rewriting 
$
 H=E+(H^d-E)\bigl(1-(E-H^d)^{-1}H^{\perp}\bigr)
$.  
A few assumptions are in need:
i) $|E\rangle$ ends up with a finite order perturbation in $H^\perp$ powers if $H^\perp$ is nilpotent on the subspace in which an $H$ eigenstate is built. A matrix $A$ is said to be nilpotent if $A^n=0$ for some positive integer $n>1$.
ii) suppose i) is satisfied, then $|E\rangle$ is constructed from a multiplet of member states ranked by the number  of powers of $H^\perp$ action ascending from a father state. We shall assume within each multiplet the $H^d$ spectrum is not degenerate for generic perturbation parameters. For simplicity we shall assume that $H^\perp$, when acts to the right,  actually maps a member state to its brother states with smaller $H^d$ eigenvalues.
iii) With the above assumptions in mind, one can show that the exact eigenstate is in fact obtained by a similarity transformation $S$ from the corresponding father state. Thus the integrability of the $H$ system inherits from that of the unperturbed $H^{(0)}$ system. In other words, any diagonal action in the unperturbed system conjugated by $S$, will remain mutually commutable  when the prescribed perturbation is turned on.
The similarity transformation $S$  is defined by the following time ordered multi-integration, 
\bea
S&=&T \exp\bigl(\int_{-\infty}^0 H^\perp (t) dt\bigr),\\\nonumber
H^\perp (t) &=&\exp(-tH^d)H^\perp\exp(t H^d).
\eea
Here $T$ means time ordering with larger $t$ to the left, and $H^\perp(-\infty)=0$  because of our convention that $H^\perp$  lowers the maximal energy of the Hilbert space when acts to the right. It can be verified that
$
S|E^{(0)}\rangle_0 =\bigl(1-(E-H^d)^{-1}H^{\perp}\bigr)^{-1}|E^{(0)}\rangle_0 =|E\rangle
$. 
 One can think of $S$\footnote{The construction of $\log(S)$ resembles that of the screening charges in 2d CFT, although in later cases the analog of $H^\perp$, which is $V_{\alpha_{\pm}}(1)$,  does not seem to be triangular.} as  an action to the right by adiabatically turning on the perturbation $H^\perp$ from time $-\infty$ to time $0$.  Consequently,
\be
S^{-1}= T \exp\bigl(-\int_0^{\infty}H^\perp (-t) dt\bigr).
\ee 
The orthogonality maintains if the conjugate state is defined by
$
_0\langle E^{(0)}|S^{-1} =_0\langle E^{(0)}|\bigl(1-H^{\perp}(E-H^d)^{-1}\bigr)^{-1} =\langle E|
$. 
One can further show that
$H=H^d+H^{\perp}=S H^dS^{-1}$. 
Thus the conditions that restrict our construction is about the same as for which the adiabatic theorem in quantum mechanics could apply. We may identify the $S$ transformation as an adiabatic transformation.
We believe the procedure should work for a class of integrable models. So in this letter we shall concentrate ourselves on the CS model ($\beta>0$ will be assumed in this letter in accordance with our convention of time ordering). The merits lying behind  this construction is that the interacting fermion system can be regarded as an adiabatic mapping by a similarity transformation from the free fermion system. This is in the same spirits as the Landau Fermi liquid theory and the 1d Luttinger liquid theory.

Being an integrable system, the CS model is exactly solvable. Though explicit constructions of the eigenstates in the second quantized form has not appeared prior to our present work. The integrability originates from the essentially free quasiparticle spectrum which accounts for the fractional statistics. This has been considered from various point of views elsewhere\cite{ Haldane:1991, Azuma:1993ra, Pasquier:1994cs, Wu:1994, Lapointe(1995), Polychronakos:1992zk}.
Although we generally work on the CS model with positive $\beta$,   negative rational values of $\beta$ for the Jack polynomials have been proposed in \cite{Bernervig:2008} to unify the FQHE wave functions of the Laughlin, More-Read and Read-Rezayi type in one picture. \cite{Estinne:2010} also relates the non-abelian statistics to the CS model through differential equations for degenerate conformal blocks. \cite{Estienne:2011qk} has gone even further by unveiling a deep connection between 2d $WA_{k-1}$ minimal models and the integrability of the generalized CS models. 

\cite{Stanley} contains a comprehensive review on the Jack symmetric function  which are the spectrum generation function for the CS model. Our recent work\cite{Wu:2011cya}, in which more relevant references can be found, also constitute a concise introduction on the subject. The CS model is introduced for studying N interacting
particles distributed on a circle of circumference $L$ with the Hamiltonian,
\begin{eqnarray}
H_{CS}&=&-\sum_{i=1}^N\frac{1}{2}\partial_{x_i}^2+\sum_{i<j}\frac{\beta(\beta-1)}{
\sin^{2} (x_{ij})}.
\end{eqnarray}
Here for convenience, we have set  $ \hbar^2/m =1$, $L=\pi$. For simplicity, we shall restrict ourselves to the following simple solutions of the eigenfunctions (for more general boundary conditions, see \cite{Doyon:2006ph,Estienne:2011qk}),
$
\Psi_\lambda(\{x_i\})=\Psi_0(\{x_i\})J_\lambda^{1/\beta}(\{z_i\})
$. 
Here, the ground state $\Psi_0(\{x_i\})$ is the Jastrow-like wave function, $\Psi_0(\{x_i\})=\prod_{i<j}\sin^\beta (x_{ij})$, $J_\lambda(\{z_i\})$ is the Jack symmetric polynomial with $z_j=\exp(2ix_j)$. 
For each Young tableau $\lambda=\{\lambda_1,\lambda_2,\cdots,\lambda_N\}$, with $\lambda_i\geq \lambda_{i+1}\geq 0$, we normalize the energy eigenvalue as $2E_\lambda$, 
$
E_\lambda=\sum P_i^2, \, \,  P_i=\lambda_i+\beta\bigl((N+1)/2-i\bigr)
$.  
It is known that the Jack polynomial is triangular in the sense that it is a linear superposition of the squeezed states starting from a dominant symmetric monomial. However, in this symmetric monomial basis, it is difficult to separate the interacting Hamiltonian to $H^{\parallel}$ and $H^{\perp}$ parts.
See however, \cite{Kadell} for diagonalization on this basis.
So we have to find  other basis in which our method could apply. For this reason we prefer to work on the second quantized form of the Hamiltonian for the collective motion of the CS model, 
\bea H &=& k\sum_{n,m>0}( a_{-n} a_{-m} a_{n+m} + a_{-n-m}a_n a_m)\\\nn &+&\sum_{n>0}\bigl(N\beta+(1-\beta) n\bigr) a_{-n} a_n.
\eea 
Here $ \beta=k^2$, 
$k$ is the charge unit of the $N$  identical particles, $[a_n,a_m]=n\delta_{n+m,0},\, [a_0,q]=1$. The ground state energy $E_0$ is no longer included in $H$. This is the bosonic picture of the CS system which describes the density fluctuation
of the electrons. To transform this Hamiltonian to the original CS Hamiltonian, we shall use the vertex operator formalism defined by
\[V_k(z_i)= \exp\bigl(k\sum_{n>0}a_{-n}z_i^{n}/n\bigr) \exp\bigl(-k\sum_{n>0}a_n z_i^{-n}/n\bigr) e^{kq}z_i^{k a_0},\] and $\Psi_\lambda(\{x_i\})=\langle k_f|J_\lambda \prod_{i=1}^{N}V_k(z_i)|k_{in}\rangle$. Here $a_0 |k_{in}\rangle =k_{in} |k_{in}\rangle, \,   k_{in} =-(N-1)k/2, \, k_f=(N+1)k/2$.  $J_\lambda$ 
solves the equation
$
\langle 0|J_\lambda H =\langle 0|J_\lambda (E_\lambda-E_0)
$.  
Then we have
$
H_{CS}\Psi_\lambda(\{x_i\})=\langle k_f|J_\lambda (H+E_0)2\prod_{i=1}^{N}V_k(z_i)|k_{in}\rangle
=2E_\lambda \Psi_\lambda(\{x_i\})
$. 
$J_\lambda$'s are the Jack symmetric functions in the power sum basis, $J_\lambda\equiv J_\lambda^{1/\beta}(\{a_n/k\})$, not in an apparently squeezed form. So in the bosonic picture, $H$ does not warrant an explicit  decomposition into $ H^{\parallel}$ and $H^{\perp}$ parts.  However, we know that for $\beta=1$, the Jack states reduce to the Schur states, which corresponds to a free non-relativistic spinless ``chiral''  fermion theory. This suggests that we may rewrite $H$ as an interacting
fermion theory with perturbation parameter $\beta-1$. 
The construction of the Schur states in the fermionic picture is made possible by the standard bosonization (for convenience we assume Neveu-Schwarz (NS) boundary condition for the moment),
$
a_{n} = \sum_{r\in\mathbb{Z}+1/2}:b_{n-r}c_{r}:
$, 
with  $\{b_r, c_s\}=\delta_{r+s,0},\, r,s\in\mathbb{Z}+1/2 $ and $b_r\ket{0}=c_r\ket{0}=0,\, r>0$.  Hereafter for simplicity we shall drop the term $\beta N\sum_{n>0} a_{-n}a_n$ in the original $H$, for it just adds a value $\beta N|\lambda|$. 
The Schur functions are the eigenstates of $H$ at $\beta=1$, 
$
H^{(0)} \equiv H_{\beta=1}=\sum_{r>0}(r^2+\frac{3}{4})(b_{-r}c_r - c_{-r}b_r),\, 
E^{(0)}_{\lambda} = \sum_{i=1}^{d(\lambda)}(r_i^2-s_i^2) = \sum_{i=1}^{\lambda^t_1}\lambda_i^2 - \sum_{i=1}^{\lambda_1}(\lambda_i^t)^2
$. 
Each Schur state is created by a monomial of equal number $d(\lambda)$ of $b_{-r}$'s and $c_{-s}$'s acting on the vacuum state with $d(\lambda)$ the number of squares along the diagonal line of $\lambda$, 
$
\ket{\lambda}\equiv s_\lambda\ket{0} = (-1)^{\sum_{i=1}^{d(\lambda)}(1/2-s_i)}\prod_{i=1}^{d(\lambda)}b_{-r_i}c_{-s_i}\ket{0}
$. 
The Schur function in the fermionic picture is labeled by the Maya diagram which is translated into the Young tableau\cite{Jimbo} this way: $r_i = \lambda_i - i +1/2, \, s_i =\lambda^t_i - i +1/2$. Here, $\lambda = \{\lambda_1,\lambda_2, \dots\}$ denotes the Young tableau and $\lambda^t = \{\lambda_1^t,\lambda^t_2,\dots\}$ its transposed Young tableau. 
For $\beta\neq 1$ CS model, the two body interaction appears and the interaction strength is proportional to $\beta -1$.  Therefore we need to eliminate any
odd powers of $k$ in $H$, which make branch cuts in the coupling space  after fermionization. This can be done by the following redefinition, 
$\tilde{a}_{-n}=a_{-n}/k,\, \tilde{a}_{n}=k a_n,\, n>0$ 
and $\tilde{a}_{0}=k a_0,\, \tilde{q}=q/k$. 
We call the above non-unitary similarity transformation the $D$ transformation, which keeps the bosonic commutators invariant, and as we shall see, also makes the Hamiltonian triangular in the fermionic picture. Making the standard bosonization to $\tilde{a}_n$'s, one found that the Hamiltonian $H$ can be written as
\bea\label{FermCS}
H &=&H^{(0)}+H^{\parallel} +H^{\perp}, \\\nonumber
H^{\parallel} &=& \sum_{r>0} (1-\beta)(r-\frac{1}{2})\bigl(\frac{1}{3} b_{-r}c_r +(r+\frac{1}{6})c_{-r}b_r\bigr)\\\nonumber &+& \sum_{r+s>0}\frac{2}{3}(2r+s)(1-\beta):b_{-s}c_{-r}b_r c_s:, \\\nonumber
H^{\perp} &=&(1-\beta)\sum_{\begin{subarray}{c}r+s>0,r+l<0\\k+l+r+s =0\end{subarray}}\bigl(2r+\frac{2}{3}(s+l)\bigr):b_k c_l b_r c_s: .
\eea 
$H^{\perp}$ is strictly triangular. That is to say,  it always squeezes the original Young tableau $\lambda$ for a given Schur state to the ``thinner'' ones $\lambda^\prime$'s for states after its action, \be\label{squeeze}\lambda^\prime<\lambda \Rightarrow \sum_{i=1}^{j} \lambda^\prime_i < \sum_{i=1}^{j} \lambda_i, \, \text{for } \, j=1,2,\cdots \, .\ee
 To see this triangular nature in a more transparent form, 
$H^\perp$ is simplified and decomposed into 5 subprocesses,
  \bea\label{triangularH}
\nn \frac{1}{2(1-\beta)}H^\perp &=&\sum_{\begin{subarray}{c}r+s>0,r+l<0\\r>k,k+l+r+s =0\end{subarray}}(r-k):b_k c_l b_r c_s: \\\nn
&=& \sum_{n=1,r>s>0}^{n=s-1/2} (s-r) c_{-r-n}c_{-s+n}b_s b_r\\\nn
&+&\sum_{n=1,r>s>0}^{n=[(r-s-1)/2]}(s-r+2n)b_{-r+n}b_{-s-n}c_s c_r\\
&+&\sum_{n=1,r,s>0}^{n=s-1/2}(r+s-n)b_{-s+n}c_{-r-n}b_r c_s\\\nn
&+& \sum_{r>l>0,s>0}(l-r) c_{-l-s-r}b_l b_r c_s\\\nn
&+&\sum_{l>r>0,s>0} (l-r) b_{-l}c_{-s}b_{-r} c_{r+s+l}\,,
\eea
here $[x]$ stands for the integer part of the number $x$. 
Each line in the above expression stands for a process of ``squeezing'' (moving downwards plaquettes in) the Young tableau representing the fermion monomial in agreement with (\ref{squeeze}).
While process 1)-3) does not change $d(\lambda)$, process 4) or 5) make it changed by $\mp 1$. 
$H^{\parallel}$ shifts the energy-eigenvalue of the Schur state from $E^{(0)}_{\lambda}$ to $E_{\lambda}^{1/\beta} = \sum_{i=1}^{\lambda^t_1}\bigl(\lambda_i^2 -\beta(2i-1)\lambda_i\bigr)$, which is the eigenenergy for the Jack state. $H^{\perp}$, however, only changes the fermion monomial (Schur state) to a fermion polynomial (Jack state) and does not change the eigenvalue. Let's first concentrate on the ket state $\ket{P^{1/\beta}_\lambda}\equiv S(k)\ket{\lambda} =D R_{\lambda}\ket{\lambda}$. Here, $D=\exp\bigl(-\log(k)(q a_0+\sum_{n>0} a_{-n}a_n/n)\bigr)$ and $R_{\lambda} = \bigl(1-(E_\lambda^{1/\beta}-H^d)^{-1}H^\perp\bigr)^{-1},\,S(1)\equiv S$ and $S(k)=DS$ has also scaled back the $1/k$ factor for the $a_{-n}$'s ($n>0$ and $a_{n}$'s will gain a factor $k$) are related to the fermionic oscillators through standard bosonization. This way $H\equiv S(k) H^d S^{-1}(k)$ will resume hermiticity (no longer triangular).
Similarly, $\bra{P_{\lambda^t}^{\beta}}\equiv\bra{\lambda} S^{-1}(k)$. 
Here we have used the duality relation $P_{\lambda^t}^{\beta}({-k a_n})\propto P_{\lambda}^{1/\beta}({a_n/k})$. The orthogonality is obvious: $\bra{P_{\chi^t}^{\beta}}P_{\lambda}^{1/\beta}\rangle = \langle\chi\ket{\lambda}=\delta_{\chi,\lambda}$. 
For a standard-normalized Jack symmetric function, $J_{\lambda}^{1/\beta}=(a_{-1}/k)^{|\lambda|}+\cdots$, we have $\bra{J_{\chi}^{1/\beta}}J_{\lambda}^{1/\beta}\rangle = \delta_{\chi,\lambda}j_\lambda $.  Here $j_{\lambda}= A_{\lambda}^{1/{\beta}}B_{\lambda}^{1/\beta}$, 
\bea
A_{\lambda}^{1/\beta}&=& \prod_{s\in
\lambda}\left(a_{\lambda}(s)\beta^{-1} +
l_{\lambda}(s)+1\right),\\\nn
B_{\lambda}^{1/\beta} &=&
\prod_{s\in\lambda}\left((a_{\lambda}(s)+1)\beta^{-1}+l_{\lambda}(s)\right).\eea 
$a_{\lambda}(s)$ and $l_{\lambda}(s)$ are called arm-length and
leg-length of the box $s$ in the Young tableau $\lambda$, 
$a_{\lambda}(s)=\lambda_i-j$, $l_{\lambda}(s)=\lambda^t_j-i$. With this normalization, we have
$|J_{\lambda}^{1/\beta}\rangle = |P_{\lambda}^{1/\beta}\rangle A_{\lambda}^{1/\beta}$ and $\bra{J_{\lambda}^{1/\beta}}= B_{\lambda}^{1/\beta}\bra{P_{\lambda^t}^{\beta}}$. 
We have checked this fermionic formalism for Jack states up to level 4, all of them match with those obtained from the known bosonic examples (solved by brute force) as desired. We now write down the level 3 results for readers' reference,
\bea\label{Result}
\nonumber\left|\right.J^{1/\beta}_{\tiny\yng(1,1,1)}\left.\right\rangle &=& 6\left|\right.{\tiny\yng(1,1,1)}\left.\right\rangle, \\\nonumber
\left|\right.J^{1/\beta}_{\tiny\yng(2,1)}\left.\right\rangle &=& \dfrac{2\beta+1}{\beta}\left|\right.{\tiny\yng(2,1)}\left.\right\rangle+\dfrac{2(\beta-1)}{\beta}\left|\right.{\tiny\yng(1,1,1)}\left.\right\rangle,
\\\nonumber \left|\right.J^{1/\beta}_{\tiny\yng(3)}\left.\right\rangle &=& \dfrac{(\beta+2)(\beta+1)}{\beta^2}\left|\right.{\tiny\yng(3)}\left.\right\rangle\\\nonumber &+&\dfrac{2(\beta-1)(\beta+1)}{\beta^2}\left|\right.{\tiny\yng(2,1)}\left.\right\rangle +\dfrac{(\beta-1)(\beta-2)}{\beta^2}\left|\right.{\tiny\yng(1,1,1)}\left.\right\rangle\,.
 \eea
 The integrability of the CS model is also nicely incorporated in our formalism. The usual Dunkl exchange operator or Sekiguchi differential operator does not apply here since there is no simple way translating the coordinate formalism to the collective mode formalism for higher order invariants. To get the CS spectrum, put $N$ vertex operators $V_k(z_i)$'s acting successively on the $|k_{in} -k/2\rangle$ vacuum starting from $V_k(z_N)$. If only the creation modes are taken into account, the resulting state is a linear superposition of the following modes labeled by Young tableau  (with maximal $N$ rows),
$V_{(1-N)\beta/2-\lambda_1}\cdots V_{(N-1)\beta/2-\lambda_N}|k_{in}-k/2\rangle$. Here the i-th mode carries the momentum $\bigl((N+1)/2-i\bigr)\beta+\lambda_i$ and the CS energy is just summing over each momentum square. Notice that for $\beta=1$,  we come back to the free fermion theory (NS sector $\Rightarrow N\in even$). In this case we can  define a momentum operator for the  specific fermionic mode $b_{-r}$, $P^{(0)}_r={\tiny{\times}\atop\tiny{\times}}r b_{-r} c_r{\tiny{\times}\atop\tiny{\times}} $. Here the normal ordering ${\tiny{\times}\atop\tiny{\times}}\cdots {\tiny{\times}\atop\tiny{\times}}$ is defined with respect to the ``empty'' vacuum  $|k_{in}-1/2\rangle$ at $\beta=1$, 
$b_r |k_{in}-1/2\rangle=0,\, r>N/2$. 
$$^{\tiny{\times}}_{\tiny{\times}}{b_r c_s}^{\tiny{\times}}_{\tiny{\times}}=\begin{cases} b_r c_s, &\text{if } r < N/2; \\-c_s b_r, &\text{if } r > N/2. \end{cases}$$  For $\beta\neq 1$ CS model, we choose to
stay in the fermionic picture, so that the canonical commutation  as well as the normal ordering just defined remains valid. To
produce the exact CS spectrum,  we just need to
define the shifted momentum operator for each fermionic mode, in a way similar to the minimal coupling of
a self-generated fictitious gauge potential. This pseudo-momentum operator for a specific mode is in fact for a collective motion, since additional information on each  electron's relative position among the total of $N$ electrons is needed,
$P^d_r\equiv P^{(0)}_r+P^{\parallel}_r={\tiny{\times}\atop\tiny{\times}}b_{-r} c_r{\tiny{\times}\atop\tiny{\times}}\bigl(r+(\beta-1)((N+1)/2-\sum_{s\geq r} {\tiny{\times}\atop\tiny{\times}}b_{-s} c_s{\tiny{\times}\atop\tiny{\times}})\bigr)$. For $\beta=1$,  $P^{\parallel}_r$ vanishes and the ``gauge'' potential drops out, and  we get exactly the momentum operator $P^{(0)}_r$ for the mode $b_{-r}$. For $\beta\neq 1$, a self-generated ``gauge'' potential has to be coupled. The ground state is specified by the null Young tableau, and the Fermi sea is filled up to momentum $(N-1)/2$. 
We call this filled Fermi sea the perturbative vacuum state $|f\rangle$, $b_r |f\rangle=0$ for $r> -N/2$. Since there are two vacuum states considered, there exists two kinds of normal ordering each associated with different vacuum state, which one to choose depends on the context. For example, in constructing the Schur or Jack state, we are doing perturbation around the filled Fermi sea $|f\rangle$, so it is better to work with the following  normal ordering $$:b_r c_s:=\begin{cases} b_r c_s, & \text{if } r < -N/2; \\-c_s b_r, & \text{if } r > -N/2.
\end{cases}$$ 
Now define
$
H^d=\sum_r (P^d_r)^2
$.  
If the i-th electron's momentum is moved  up exactly by $\lambda_i$  amount, then  $H^d$ acts on this system will produce the exact CS spectrum.  $S(k)$ act on this fermion monomial state will produce the exact Jack state. Since  $
[P^d_r, P^d_s]=0 $, the conserved charges can now be constructed, 
$W^{n}=S(k)\sum_r (P^d_r)^n S^{-1}(k)\Rightarrow [W^n,W^m]=0,\, n,m>0$.

We have shown in this letter that Jack symmetric function
is triangular in the basis of Schur functions. On the other hand, expanding in symmetric monomial basis $m_\mu$, $P_\lambda^{1/\beta}(\{z_i^n\})=(v^{1/\beta})_\lambda^\mu m_\mu=R_\lambda^\nu s_\nu(\{z_i^n\})\Rightarrow R_\lambda^\nu=(v^{1/\beta})_\lambda^\mu ((v^{1})^{-1})_\mu^\nu$. Here, $\{(v^{1/\beta})_\lambda^\mu \}$ as well as $\{((v^{1})^{-1})_\mu^\nu\}$ is a triangular matrix with unit diagonal entry. This shows again that the matrix $\{R_\lambda^\nu\}$, as well as $H$, with their explicit form given in this letter, is  triangular. 
To get the bosonic formalism  we can use the well-known Frobenius formula to expand the Schur polynomial in the basis of power-sum polynomials and the transition coefficient is proportional to the character for the related representation evaluated in the conjugacy class of symmetric group\cite{Lassalle}. 

This work is part of the CAS program ``Frontier Topics in Mathematical Physics'' (KJCX3-SYW-S03) and
is supported in part by a national grant  NSFC(11035008).

\end{document}